\documentclass[12pt]{article}
\usepackage{epsfig}
\begin{document}
\begin{titlepage}
\begin{center}
\vspace{2cm}

{\Large \bf Topological charge screening and the infrared behavior of the instanton density}
\vspace{0.50cm}\\

Nikolai Kochelev$^{a,b}$\footnote{kochelev@theor.jinr.ru} \vspace{0.50cm}\\
{(a) \it School of Physics and Astronomy, Seoul National University,\\ Seoul 151-747,  Korea}\\
{(b) \it Bogoliubov Laboratory of Theoretical Physics, Joint
Institute for Nuclear Research, Dubna, Moscow region, 141980
Russia} \vskip 1ex
\end{center}
 \vskip 0.5cm \centerline{\bf Abstract} A new
mechanism for suppression of the instanton density in the infrared is
considered. This mechanism is based on the phenomenon of topological charge
screening, which leads to an effective cutoff in the contribution of
large instantons. \vspace{1cm}

\end{titlepage}

\setcounter{footnote}{0}

Instantons, strong topological fluctuations of gluon fields in
QCD, are widely believed to play an important role in the physics
of the strong interaction (for reviews see \cite{shuryak},
\cite{diakonov}). In particular, instantons provide mechanisms for
the violation of both $U(1)_A$ and chiral symmetry in QCD, and may
therefore be important in determining hadron masses and in the
resolution of the famous $U(1)_A$ problem. Furthermore, it was
recently shown that instantons persist through the deconfinement
transition, so that instanton-induced interactions between quarks
and gluons may underlie the unusual properties of the so-called
strongly coupled quark-gluon plasma recently discovered at RHIC
\cite{shuryak1}.

One longstanding problem in the calculation of instanton
contributions to physical processes is the uncertainty
in their distribution over the full range of instanton
length scales. For small instantons in the instanton
dilute gas approximation this distribution is well known,
\begin{equation}
D_0(\rho)=\frac{d}{\rho^5}\Bigl(\frac{2\pi}{\alpha_s(\rho)}\Bigr)^{2N_c}
e^{-S(\rho)} \label{gas}
\end{equation}
where $d$ is a renormalization scheme dependent constant, and
\begin{equation}
S(\rho)=S_0(\rho)=\frac{2\pi}{\alpha_s(\rho)} \label{action}
\end{equation}
is the Euclidean action of the instanton solution. This distribution
increases rapidly with increasing instanton size, as
$D_0(\rho)\approx \rho^{\beta_0-5}$ (where $\beta_0=11N_c/3-2N_{f}/3$).
The resulting instanton density integral is divergent, so
it is impossible to make definite quantitative predictions regarding
instanton effects given this distribution. Shifman, Vainshtein and Zakharov
(SVZ) \cite{SVZ} have shown that at an instanton length scale of
$\rho\approx 1\; {\rm GeV}^{-1}$ the interaction between instantons and
the gluon condensate becomes strong, which leads to a significant
change in the instanton density at large instanton length scales,
\begin{equation}
D_{cond}(\rho)\approx
D(\rho)_0 \exp\bigg\{\frac{\pi^4\rho^4}{8\alpha_s^2(\rho)}
<0|\frac{\alpha_s}{\pi}\, G^a_{\mu\nu}G^a_{\mu\nu}|0>\bigg\}
\label{cond}
\end{equation}
where the gluon condensate is numerically \cite{ioffe}
\begin{equation}
<0|\frac{\alpha_s}{\pi}\, G^a_{\mu\nu}G^a_{\mu\nu}|0> = 0.009\pm 0.007
\;{\rm GeV}^4 \label{g2}.
\end{equation}
Evidently this interaction leads to an even stronger divergence in
the instanton density at large length scales than in
Eq.(\ref{gas}). Recent lattice calculations however show a damping
in the rate of increase of the instanton density for instanton
lengths of $\rho_c\sim 0.3-0.5 $~fm, followed by a rapid decrease
at larger instanton sizes \cite{lattice}. Several scenarios have
been suggested to explain the physics behind this phenomenon. One
possible mechanism for this decrease is a repulsive interaction
between large instantons. This idea was suggested by Ilgenfritz
and M\"uller-Preussker in a papers \cite{ilgenfritz} and was
developed by Diakonov and Petrov  later on \cite{petrov}. The
effect of confinement on the instanton distribution,  considered
by Shuryak, is another possible mechanism for the suppression of
the instanton density at large $\rho$ \cite{shuryak2}. In a paper
by Dorokhov {\it et al.} \cite{dorokhov} it was shown that the
deformation of instantons in external vacuum fields might also
lead to a suppression of the density of large instantons.

In this Letter we consider another possible mechanism for the
suppression of large instantons, which is the influence
of topological charge screening on instanton density.
This screening effect is known to play an important role in the
properties of the $\eta^\prime$ meson, and leads in particular
to a vanishing topological susceptibility
\begin{equation}
\chi(k^2)=i\int d^4x\, e^{ikx}<0|T\{Q_5(x)Q_5(0)\}|0>
\label{top}
\end{equation}
as $k^2\to 0$, in the chiral limit \cite{shuryak}.
In Eq.(\ref{top}),
\begin{equation}
Q_5(x)=\frac{\alpha_s}{8\pi}\, G^a_{\mu\nu}(x)\widetilde{G}^a_{\mu\nu}(x)
\label{topd}
\end{equation}
is the topological charge density, and
$\widetilde{G}^a_{\mu\nu}=1/2\epsilon_{\mu\nu\alpha\beta}
G^a_{\alpha\beta}$ is the dual gluon field strength tensor.

Our starting point is the SVZ interaction between an instanton
and an external gluon field \cite{SVZ}:
\begin{eqnarray}
{\cal L}_{eff}(x_0)&=&\int d\rho \, D_0(\rho)
\exp\bigg({-\frac{2\pi^2\rho^2}{g_s}\, \bar\eta_{a\alpha\beta}G^a_{\alpha\beta}(x_0)}
\bigg).
\label{lag}
\end{eqnarray}
Here $\bar\eta_{a\alpha\beta}$ is 't Hooft's symbol, and $x_0$ is the
instanton's location.
On expanding Eq.(\ref{lag}) in powers of the gluon field
strength $G_{\alpha\beta}$, and keeping only terms with an even number of gluon
fields, we find
 \begin{eqnarray}
{\cal L}_{eff}&=&\int d\rho \,
D_0(\rho)\Big[1+\frac{1}{2!}\Bigl(\frac{2\pi^2\rho_c^2}{g_s}\Bigr)^2\bar\eta_{a\alpha\beta}
\bar\eta_{b\mu\nu}
 G^a_{\alpha\beta}G^b_{\mu\nu}\nonumber\\
&+&\frac{1}{4!}\Bigl(\frac{2\pi^2\rho_c^2}{g_s}\Bigr)^4\bar\eta_{a\alpha\beta}
\bar\eta_{b\mu\nu}\bar\eta_{c\rho\sigma}\bar\eta_{f\tau\gamma}
 G^a_{\alpha\beta}G^b_{\mu\nu}G^c_{\rho\sigma}G^f_{\tau\gamma}+...\Big].
\label{lag2}
\end{eqnarray}
The effect of the gluon condensate on the instanton density at leading order is
\begin{equation}
{\cal L}_{eff}=\int d\rho \,
D_0(\rho)\Big[1+\frac{\pi^4\rho^4}{8\alpha_s^2(\rho)}
<0|\frac{\alpha_s}{\pi}\, G^a_{\mu\nu}G^a_{\mu\nu}|0>+...\Big],
\end{equation}
which follows from the second term in Eq.(\ref{lag2}),
after using the color-singlet nature of the vacuum state
\begin{equation}
<0|G^a_{\alpha\beta}G^b_{\mu\nu}|0>=\frac{1}{N_c^2-1}
\delta^{ab}<0|G^a_{\alpha\beta}G^a_{\mu\nu}|0> \label{avar}
\end{equation}
and an identity for 't Hooft symbols in Minkowski space-time,
\begin{equation}
\bar\eta_{a\alpha\beta}\bar\eta_{a\mu\nu}
=g_{\alpha\mu}g_{\beta\nu}-g_{\alpha\nu}g_{\beta\mu}-i\epsilon_{\alpha\beta\mu\nu}.
\label{sum}
\end{equation}
The expansion of Eq.(\ref{lag}) to high orders, combined with an
assumption of factorization for the higher $<0|G^{2n}|0>$ condensates,
leads to the exponential form given in Eq.(\ref{cond}).

We emphasize that in their original derivation of the effect of gluon condensates
on the instanton density \cite{SVZ}, SVZ ignored all terms involving
the dual tensor $\widetilde{G}^a_{\mu\nu}$. Actually only the leading term in the
dual tensor expansion vanishes,
\begin{equation}
<0|Q_5(x)|0>=0,
\label{vac}
\end{equation}
due to the $CP$ invariance of the QCD vacuum.
Terms involving higher powers of $\widetilde{G}^a_{\mu\nu}$
generally do contribute to the expansion in Eq.(\ref{lag2}).
The leading dual tensor contribution arises from the third term
in Eq.(\ref{lag2}), and involves the matrix element
\begin{equation}
M=<0|\bar\eta_{a\alpha\beta}
\bar\eta_{b\mu\nu}\bar\eta_{c\rho\sigma}\bar\eta_{f\tau\gamma}
 G^a_{\alpha\beta}G^b_{\mu\nu}G^c_{\rho\sigma}G^f_{\tau\gamma}|0>.
\end{equation}
By using the formula for the vacuum matrix element
\begin{equation}
<0|G^a_{\alpha\beta}G^b_{\mu\nu}G^c_{\rho\sigma}G^f_{\tau\gamma}|0>\rightarrow
\frac{\delta^{ab}\delta^{cf}}{N_c^4-1}
<0|(G^d_{\alpha\beta}G^d_{\mu\nu}G^m_{\rho\sigma}G^m_{\tau\gamma}+perms.)|0>,
\label{col}
\end{equation}
one obtains
\begin{equation}
\Delta {\cal L}_{eff}\sim\int d\rho \,
D_0(\rho)\frac{\pi^6\rho^8}{5\cdot 2^5\alpha_s^4(\rho)}
<0|\alpha_sG^a_{\mu\nu}(x_1)\widetilde{G}^a_{\mu\nu}(x_2)\alpha_sG^b_{\alpha\beta}(x_3)
\widetilde{G}^b_{\alpha\beta}(x_4)|0>,
\label{ff}
\end{equation}
where $x_i\approx x_0$. In Eq.(\ref{ff}) the nonlocality of the instanton
is explicitly taken into account.
This differs from the
contribution of a pure gluon condensate, for which one can use
the local approximation $x_i=x_0$. The difference follows from
the singular behavior of the topological density at the origin.
To illustrate this, note that in the instanton model of the QCD vacuum
the correlator of topological density in the chiral limit is \cite{shuryak}
\begin{equation}
<0|Q_5(x)Q_5(0)|0>\approx
\frac{N}{V}\Big[\delta^4(x)-\frac{2N_f}{f_\pi^2}
\frac{N}{V}<0|\eta^\prime(x)\eta^\prime(0)|0>\Big],
\label{screen}
\end{equation}
where $N/V$ is the instanton density. The local term in
Eq.(\ref{screen}) gives the contribution of a single instanton
centered at the origin, and the second term is the so-called
screening contribution. Evidently one should ignore the
contribution of the first term in Eq.(\ref{screen}) in estimating
the matrix element in Eq.(\ref{ff}), because it arises from the instanton
under consideration. Therefore in the limit $|x_i-x_j|\to 0$ we obtain
\begin{equation}
<0|\alpha_sG^a_{\mu\nu}\widetilde{G}^a_{\mu\nu}\alpha_s
G^b_{\alpha\beta}\widetilde{G}^b_{\alpha\beta}|0>^{scr} \approx
-\frac{4}{\pi^2} \int d^4k_E\, \bar\chi(k^2_E),
\label{sc2}
\end{equation}
where $\bar\chi(k^2_E)$ is the screening contribution to the
topological susceptibility (for Euclidean momenta). It should be
noted that the exact form of $\chi(k^2)$ is unknown, and only
model calculations of this function are available (see \cite{shuryak},
\cite{ioffe2}, \cite{dor2}). Here we are interested in the long
range behavior of $\chi$, which can be represented by the
expression
\begin{equation}
\chi(k^2_E)\approx
-\chi^\prime(0)\, m^2_{\eta^\prime}\, (1-e^{-k^2_E/m_{\eta^\prime}^2}),
\label{ch1}
\end{equation}
where the value of the derivative $\chi^\prime(0)$
is known from a QCD sum rule calculation \cite{ioffe2},
$\chi^\prime(0)=(2.3\pm 0.6)\times 10^{-3}\; {\rm GeV}^2$.
From Eq.(\ref{ch1}), the screening part is
\begin{equation}
\bar\chi(k^2_E)\approx
\chi^\prime(0)\, m^2_{\eta^\prime}\, e^{-k_E^2/m_{\eta^\prime}^2}.
\label{ch2}
\end{equation}
Thus we find the estimate
\begin{equation}
<0|\alpha_sG^a_{\mu\nu}\widetilde{G}^a_{\mu\nu}
\alpha_sG^b_{\alpha\beta}\widetilde{G}^b_{\alpha\beta}|0>^{scr}
\approx -4\chi^\prime(0)\, m^6_{\eta^\prime}.
 \label{ch3}
\end{equation}
It is easy to show that the higher order expansion in
$\widetilde{G}^a_{\mu\nu}$ can be exponentiated, given the
assumption of vacuum state dominance. Our final result for the
screening correction to the instanton density is thus
\begin{equation}
 \Delta {\cal L}_{eff}\approx \int d\rho \,
D_0(\rho)\exp\bigg(-\frac{\pi^6\rho^8}{40\alpha_s^4(\rho)}
\, \chi^\prime(0)\, m^6_{\eta^\prime}\bigg).
 \label{ff1}
\end{equation}

In contrast to the $\rho^4$ dependence in the exponential factor
in Eq.(\ref{cond}), our novel correction evidently has a faster
$\rho^8$ behavior in the exponent. The sign of this contribution
is very important; since it is negative, this effect leads to a
{\it suppression} of the instanton density at large $\rho$. We
emphasize that this sign is determined by the sign of the
screening correction in Eq.(\ref{screen}), and follows from the
general property of the positivity of state norms in Euclidean
field theory \cite{osterwalder}.\footnote{We are grateful to A.Di
Giacomo for clarifying this point.}
\begin{figure}[htb]
\centering
\centerline{\epsfig{file=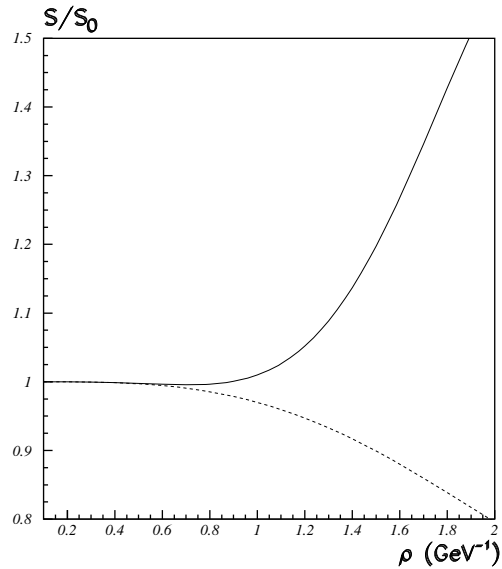,width=8cm,angle=0}}\
\caption{The ratio of our effective action to the free instanton
action, after incorporating the gluon condensate contribution (dashed)
and the screening correction (solid).}
\end{figure}

In Fig.1 we show the effective action ratio
$S_{eff}(\rho)^{cond,scr}/S_0(\rho)$,
where
$S_{eff}(\rho)^{cond}$ is the effective action including the
gluonic condensate correction, Eq.(\ref{cond}), and
$S_{eff}(\rho)^{scr}$ is the action including the
screening contribution, Eq.(\ref{ff1}).
To generate this figure the expression
\begin{equation}
\alpha_s(\rho)=\frac{4\pi}{9\ln[(1/(\rho^2\Lambda^2)]}
\end{equation}
was used to estimate the strong coupling constant,
where $\Lambda\approx 300$ MeV. It is evident in Fig.1
that the effect of screening on the instanton density becomes
dominant at rather small instanton length scales
($\rho>1$ GeV$^{-1}$), and leads to a rapid increase in the
effective instanton action at large instanton scales. One can
therefore anticipate a corresponding cutoff in the instanton
density relative to the dilute gas approximation of Eq.(\ref{gas}).

In summary, we have shown that topological charge screening
strongly affects the length scale dependence of the
distribution of instantons, and provides a natural infrared
cutoff in the instanton density.

\eject

We are happy to acknowledge useful discussions of various aspects
of this work with A.Di Giacomo, A.E.Dorokhov, E.-M.Ilgenfritz,
B.L.Ioffe, E.V.Shuryak and V.I.Zakharov. We would also like to
thank T.Barnes for proofreading the manuscript. The author is very
grateful to the School of Physics and Astronomy of Seoul National
University, and especially Prof. Dong-Pil Min, for their kind
hospitality. This work was supported in part by the Brain Pool
program of the Korea Research Foundation through KOFST grant
042T--1--1, by the Russian Foundation for Basic Research through
grant RFBR--04--02--16445.


\begin{thebibliography}{99}

\bibitem{shuryak} T. Sch\"afer and E.V. Shuryak,
Rev. Mod. Phys. {\bf 70} (1998) 1323.

\bibitem{diakonov} D.~Diakonov,
Prog.\ Part.\ Nucl.\ Phys.\  {\bf 51} (2003) 173.

\bibitem{shuryak1} E.~V.~Shuryak,
arXiv:hep-ph/0608177.

\bibitem{SVZ}
M.A. Shifman, A.I.Vanshtein and V.I. Zakharov,
Nucl.\ Phys.\ {\bf B165} (1980) 45.




\bibitem{ioffe}
B.~L.~Ioffe,
Prog.\ Part.\ Nucl.\ Phys.\  {\bf 56} (2006) 232.


\bibitem{lattice}
B.~Lucini, M.~Teper and U.~Wenger, Nucl.\ Phys.\  {\bf B715}
(2005) 461; A.~Hasenfratz and C.~Nieter, Phys.\ Lett.\  {\bf B439}
(1998) 366;

\bibitem{ilgenfritz} E.-M. Ilgenfritz and M. Muller-Preussker
  Nucl. Phys. {\bf B184} (1981) 443; Z.Phys. {\bf C16} (1983) 339,
   Erratum-ibid. {\bf C20} (1983) 186.

\bibitem{petrov} D.I. Diakonov and V.Y. Petrov, Nucl. Phys. {\bf
B245} (1984) 259.


\bibitem{shuryak2} E.V. Shuryak, hep-ph/9909458;
hep-th/0605219.

\bibitem{dorokhov}
A.~E.~Dorokhov, S.~V.~Esaibegian, A.~E.~Maximov and S.~V.~Mikhailov,
Eur.\ Phys.\ J.\  {\bf C13} (2000) 331.

\bibitem{ioffe2}
B.~L.~Ioffe,
Surveys High Energ.\ Phys.\  {\bf 14} (1999) 89;
B.~L.~Ioffe and A.~V.~Samsonov,
Phys.\ Atom.\ Nucl.\  {\bf 63} (2000) 1448
[Yad.\ Fiz.\  {\bf 63} (2000) 1527]; B.~V.~Geshkenbein and B.~L.~Ioffe,
Nucl.\ Phys.\  {\bf B166}, 340 (1980).

\bibitem{dor2}A.~E.~Dorokhov and W.~Broniowski,
Eur.\ Phys.\ J.\  {\bf C32} (2003) 79.

\bibitem{osterwalder}
K.~Osterwalder and R.~Schrader,
Commun.\ Math.\ Phys.\  {\bf 31} (1973) 83;
{\bf 42} (1975) 281.

\end{thebibliography}
\end{document}